\documentstyle[epsf,psfig,subfigure]{l-aa}

\begin{document}

\thesaurus{08 (08.19.4);	
	   11 (11.05.2;		
	      11.06.1;		
	      11.06.2);		
	   12 (12.03.4)		
          }

\title{Faint galaxies in semi--analytic models~: how robust are the 
predictions~?}

\author {
Catarina~Lobo \inst{1,2}
\and
Bruno~Guiderdoni \inst{1}
}

\offprints{C.~Lobo, lobo@brera.mi.astro.it}

\institute{
	Institut d'Astrophysique de Paris, CNRS,
        98bis Bd Arago, F-75014 Paris, France 
\and
	Osservatorio Astronomico di Brera, via Brera 28,
	I-20121 Milano, Italy
}

\date{Received ; accepted }

\maketitle
\markboth{Lobo \& Guiderdoni}{Modelling surface brightness thresholds on 
galaxy luminosity functions}

\begin{abstract}
In spite of their overall success, semi-analytic models of galaxy formation 
and evolution predict slopes of luminosity functions which are steeper than
the observed ones. This discrepancy has generally been explained by
subtle surface brightness effects acting on the observational samples. 
In this paper, we explicitly implement the computation of surface brightness 
in a simple semi--analytic model (with standard CDM),
and we estimate the effect of 
observational surface brightness thresholds on the predicted luminosity 
functions.
The crucial free parameter in this computation is the efficiency 
$\epsilon$ of supernova feedback which is responsible for the triggering of
galactic winds. With the classical formalism for this process, 
it is difficult to reproduce simultaneously 
the Tully--Fisher relation and the flat slope of the observational luminosity 
function with the same value of $\epsilon$. 
This suggests that the triggering of galactic winds is a complex phenomenon. 
The highly uncertain formalism for supernova feedback that is used by 
semi--analytic models produces large uncertainties in the results.
However, once a value of $\epsilon$ has been chosen, the various
luminosity functions observed in different wavebands ($B$, $r$, $K$) 
and at different surface brightness thresholds, are consistently
reproduced with the surface brightness thresholds quoted by the observers.
This seems to show that these observations do see subsamples of the same 
underlying populations of ``sub--$L_\star$'' and dwarf galaxies. 
The conclusion of this heuristic paper is that a more realistic description
of SN feedback is needed, and that surface brightness effects should not be 
neglected in the modelling of galaxy formation.

\end{abstract}

\section{Introduction}\label{introducao}

Significant progress has been recently achieved in the modelling 
of galaxy formation and evolution through the development of the so--called
``{\it ab initio} semi--analytic models'' in which the astrophysics of 
gas collapse,
star formation/evolution and stellar feedback can be explicitly implemented
in the hierarchical clustering paradigm (White $\&$ Frenk 1991; Lacey $\&$ 
Silk 1991; Lacey {\it et al.} 1993, Kauffmann {\it et al.} 1993, 1994; Cole
{\it et al.} 1994; Heyl {\it et al.} 1995; Kauffmann 1995, 1996; Baugh 
{\it et al.} 1996a,b, 1997; Somerville $\&$ Primack 1998).
These models have been successfully applied to the 
prediction of the statistical properties of galaxies in the UV, visible
and NIR emitted by stars, and, very recently, in the FIR emitted by dust
(Guiderdoni {\it et al.} 1998). 
In spite of differences in the details, these studies 
lead to conclusions which are remarkably similar.

However, very little theoretical work has been done so far to implement 
observational biases in theoretical models of galaxy formation and evolution. 
Among the various possibilities, surface brightness limits are surely the most 
important bias, and this is why a detailed study of this specific effect in 
the framework of 
semi--analytic models is highly desirable. Namely, one would expect that 
introducing a surface brightness threshold into the calculation would 
drastically change the predictions issued for the luminosity functions.

As a matter of fact, this kind of improvement seems to be necessary 
to solve the issue of the luminosity function.
All the observational works concerning field galaxies give luminosity function 
slopes that are clearly 
less steep than the predicted value $\alpha \sim -1.7$,
although this latter slope already takes into account the astrophysics
of star formation/evolution and is significantly shallower than
the underlying halo mass distribution with $\alpha \sim -2$.
Two different explanations have thus been put forward to account for this 
mismatch, namely, merging of galaxies or failure of detection due to 
surface brightness limitations. 

(1) Merging of galaxies. Merging could reduce the number of objects in 
the local universe. This hypothesis has been modelled and tested by 
Kauffmann {\it et al.} (1993, 1994), Cole {\it et al.} (1994), 
Somerville $\&$ Primack (1998), 
and other papers in these series. The results seem to indicate that 
the merging rate is relatively small and that mass and luminosity 
functions issued from these models are not significantly different from 
those computed with less elaborate formalisms. 

(2) Failure of detection due to surface brightness limitations. Observational 
selection effects would provide a simple explanation 
that could interplay with the strong intrinsic dimming 
of some galaxies along their lives (see {\it e.g.} 
McGaugh 1994; Schade $\&$ Ferguson 1994; Phillipps $\&$ 
Driver 1995; Driver $\&$ Phillipps 1996; Bothun, Impey $\&$ McGaugh 
1997). For instance, a strong episode of star formation 
followed by mass loss due to supernova--driven winds and the aging of 
the remaining stellar population would cause this dimming and could even 
eventually lead to the extreme case of gravitational 
disruption, especially for dwarf 
or extended galaxies, where the potential well is shallower
(Broadhurst {\it et al.} 1988, Lacey $\&$ Silk 1991, 
Cowie {\it et al.} 1991, Babul $\&$ Rees 1992). This 
kind of scenarios would also provide a plausible explanation for the 
differences between field and cluster galaxy luminosity functions, these 
latter ones showing typically steeper faint-end slopes. As suggested by 
{\it e.g.} Babul $\&$ Rees (1992), mass loss suffered by galaxies within 
clusters is actually minimized thanks to the confinement 
provided by the surrounding intra-cluster gas. In fact, 
while steep slopes $\alpha \sim -1.8$ of luminosity functions have lately 
begun to be unveiled by new observations of clusters ({\it e.g.} Driver 
{\it et al.} 1994; De Propris {\it et al.} 1995; Lobo {\it et al.} 1997; 
Wilson {\it et al.} 1997; Trentham 1997a,b, 1998a,b), it is not 
clear yet whether field galaxy luminosity functions in the local universe 
follow the same trend. 
This is because faint and diffuse objects are marginally detected in the 
standard spectroscopic surveys of field galaxies and, for those that are 
flagged, the signal-to-noise ratio is often quite weak.

The nature of these galaxies that fail to be detected by local field 
surveys is extensively discussed in recent literature 
(see {\it e.g.} McGaugh 1995; Bothun, Impey $\&$ McGaugh 1997) and they span a 
large range in mass~: from low-mass faint dwarfs 
till the most massive, low surface brightness, blue disks ({\it e.g.} 
Malin~1~; Bothun {\it et al.} 1987).
These objects do exist in large numbers, even though they 
contribute somewhat modestly (by 10 to 30$\%$) to the total integrated 
luminosity (McGaugh $\&$ Bothun 1994, Impey {\it et al.} 1996, McGaugh 1996). 
Thus, systematic errors undoubtedly affect the 
field luminosity functions that have been published up to now, since none of 
them has had a completeness correction to account for this bias 
(Impey {\it et al.} 1996). Once the surface brightness selection effects 
are taken into account, the faint slope of the observed local field luminosity 
function is bound to suffer a considerable change (McGaugh {\it et al.} 1995). 
Computations by Driver $\&$ Phillipps (1996) predict an upturn 
in the luminosity function at $\cal{M}$$_B$~$>$~$-$18~\footnote{All 
values presented throughout this paper 
have been converted to $H_0 = 50$ km~s$^{-1}$Mpc$^{-1}$.} that has 
actually been detected by the works of Schade $\&$ Ferguson (1994), 
Marzke {\it et al.} (1994a,b) and Zucca {\it et al.} (1997). 
The works of Sprayberry {\it et al.} (1997) and Loveday 
(1997) also seem to indicate that the local field luminosity function 
definitely contains more faint members than had been detected up to recently 
by {\it e.g.} Loveday {\it et al.} 
(1992), Ellis {\it et al.} (1996), and Lin {\it et al.} (1996).

It would thus seem that a fading scenario is the most promising approach to 
tackle the disagreement between theories and observations. In the following,
we revisit this problem and analyze the effects of 
observational biases by comparing the results of galaxy modelling and
observations. More specifically, by adopting a scenario of surface brightness 
dimming, we aim at quantifying the influence of pre--determined 
surface brightness thresholds on the slope of the multi--wavelength 
luminosity function of field galaxies.

This issue leads us immediately to consider the delicate modelling of mass 
loss due to
supernova--driven winds. All semi--analytic models use the same ideas
to implement this process, originating in the pioneer paper by 
Dekel \& Silk (1986). Type II/Ib supernovae eject kinetic and thermal energy 
in the interstellar medium. Only a fraction $\epsilon$ effectively heats up 
the gas. When the thermal velocity of the gas reaches the escape velocity 
$v_{esc}$, the interstellar medium is lost through galactic winds, and 
star formation is quenched. The first problem with this standard modelling 
is that the process depends on the depth of the potential well. In principle, 
the latter is computable e.g. in CDM cosmologies. But the question is 
whether the theoretical density profiles actually reproduce the true 
density profiles of haloes. Second,
radiative losses are important and, consequently, the efficiency 
$\epsilon$ should be weak (Thornton {\it et al.} 1998). Finally, 
this model does not describe what actually occurs in disk galaxies: when
the expanding shell reaches the edge of the disk along the $z$--axis 
perpendicular to the plane, it simply blows its hot gas out, with little 
effect on the cold gas of the disk. As a matter of fact, energetic outflows 
observed in irregular galaxies seem to 
blow out only a small fraction of the insterstellar medium (Martin 1996).

When a surface--brightness criterion is taken into 
account, the strong uncertainty in the modelling of supernova feedback appears 
more clearly. Then we can test the robustness of the results at the faint
end of the luminosity function. Hereafter, we take the
standard model which is used by all authors of semi--analytic models
(see the above--mentionned 
references), and we show how the efficiency parameter $\epsilon$ (taken
as a ``fudge factor'') influences the results. Of course, a still larger 
uncertainty in the model (which appears plausible) would affect the results 
to a larger extent.

To achieve this heuristic work, we do not need a complicated cosmological 
model. We hereafter use a simple version of a semi--analytic model of 
galaxy formation and evolution, with standard CDM (the most classical case).
The layout of this paper is the following~: in section~\ref{modelo} we rapidly 
describe the model, emphasizing the new components we introduce with this 
work. These determine the behavior of the output quantities relatively to the 
star formation histories, wind efficiency and surface brightness thresholds, 
as discussed in section~\ref{fit_parametros}. Synthetic luminosity functions
are compared with observational data for redshift $z = 0$ in 
section~\ref{funcoes_luminosidade}. Finally, in section~\ref{conclusoes}, we 
discuss the results and present our conclusions.
In particular, the shape of the luminosity function 
provides us with a powerful tool for
emphasizing the strong influence of supernova feedback on the detection of 
faint galaxies when surface brightness criteria are taken into account.

\begin{figure}[htbp]
\hspace{1.cm}
\epsfysize=9cm
\epsfbox{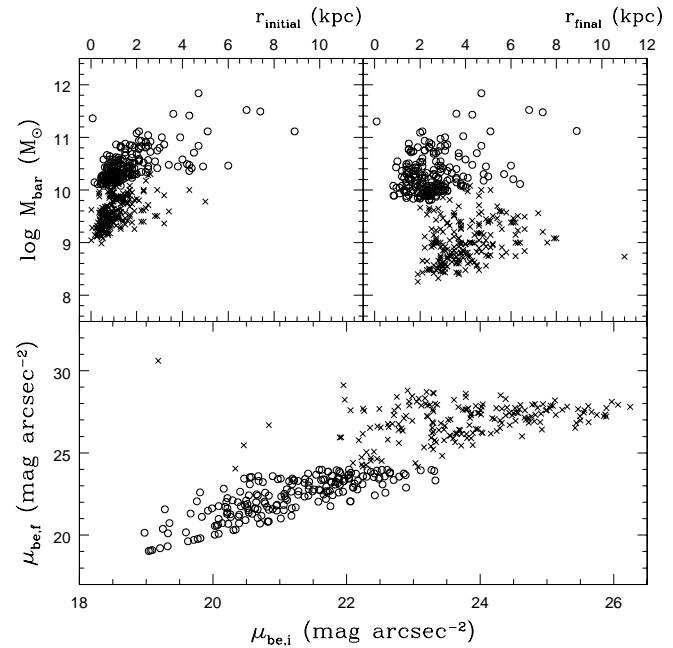}
\caption{Figure showing the effect of winds upon the radius of each galaxy, 
its baryonic mass $M_{bar}$ and its surface brightness (measured at the 
effective radius in the $b_J$ band). 
Open circles represent objects that are brighter at $z=0$ than the critical 
magnitude and surface brightness thresholds (set by observations) and 
inversely for crosses. The fudge parameter for supernova feedback is 
$\epsilon=0.8$, and the surface--brightness threshold is $\mu_{b_J}=24$ 
mag arcsec$^{-2}$.
In the upper panels, {\it initial} denotes the epoch just prior 
to star formation ignition; at that time, the baryonic mass is taken as all 
the initial 
gas mass available to form stars. Subscript {\it final} corresponds to the
present time ($z=0$), when the baryonic matter is the stellar content plus the 
remaining gas. In the lower panel, $f$ is for {\it final}, while 
$\mu_{be,i}$ denotes the surface brightness that the galaxy 
would have without radial expansion.}
\label{fig:expansao}
\end{figure}

\section{The model}\label{modelo}

We use a simple version of a semi--analytic model of formation and 
evolution of galaxies, 
to which we add some original ingredients which are relevant for our
study. In this section we will sketch 
an overview of the underlying model, leaving the detailed descriptions for the 
new modules introduced by this work. We adopt the standard Cold Dark Matter 
(SCDM) scenario -- namely, a density parameter $\Omega_0=1$, a baryon 
fraction $\Omega_{baryon}=0.05$, a cosmological constant $\Lambda=0$, a 
Hubble 
constant $H_0=50$ km s$^{-1}$ Mpc$^{-1}$, and a normalization for the power 
spectrum of linear fluctuations that is given by the inverse of the linear
bias parameter $b=1.6$ -- and we focus on the astrophysics 
of galaxy evolution. The influence of the cosmological parameters on the 
predictions of semi-analytic models is studied by {\it e.g.} Heyl {\it et al.} 
(1995) and Somerville $\&$ Primack (1998), and turns out to be smaller than 
the uncertainties due to the free parameters of the dissipative astrophysics 
(mainly star formation and stellar feedback).

\subsection{Dark matter halos}\label{modelo_DMhalos}

Simplifying assumptions need to be made in a semi--analytic formalism. In 
particular, the initial linear density perturbation is assumed to be spherical 
and homogeneous. This is the so--called ``top--hat'' model, which can be 
entirely defined by two parameters~: the size $R$ and density 
contrast $(\delta \rho /\rho)_{z=0} \equiv \delta_0$, which are the 
linearly--extrapolated values
at $z=0$, or, equivalently, by the mass $M$ and collapse redshift 
$z_{coll}$. If $\rho_0$ is the current mass density of the universe, we have 
$M = (4\pi/3) R^3 \rho_0$ and $\delta_0=\delta_{0c}(1+z_{coll})$ with 
$\delta_{0c}=1.68$ (for $\Omega_0=1$). 
The perturbation thus generated grows, 
reaches turnaround radius, decouples from the expansion of the Universe and 
collapses to form a spherical, virialized halo truncated at ``virial radius'' 
$r_V = 3R/10 \delta_0$.
 
The properties of the virialized halo are entirely computable as functions
of $M$ and $z_{coll}$. Hereafter we adopt the universal density profile 
proposed by Navarro, Frenk $\&$ White (1997, hereafter NFW) instead of 
the $r^{-2}$  ``isothermal'' profile, since the former is now used by the 
most recent versions of the semi--analytic models. Our purpose is to show the 
uncertainties in the standard situation, rather than to address the issue 
of the ``true'' density profile. Anyhow, we checked that the use of
``isothermal'' profiles would not change our conclusions qualitatively.
In the NFW model, the density profile of 
the relaxed halo at $r \leq r_V$ is~:

\begin{equation}
\frac{\rho_H(r)}{\rho_{crit}} \, = \, \frac{\delta_{char}}{(r/r_s)(1+r/r_s)^2}
\end{equation}

\noindent where $r_s$ is a scale radius, and $\rho_{crit} = 3H^2/8 \pi G$ 
is the critical density for closure. 
The characteristic density $\delta_{char}$ is computed by:

 \begin{equation}
 \delta_{char} (M,z_{coll}) \sim 3 \times 10^3 \, \Omega(z_{coll}) 
 \biggl({1+z_{coll,sub} \over 1+z_{coll}}\biggr)^3
 \end{equation}
  
\noindent where $z_{coll,sub}$ is computed according to the procedure described
in Navarro, Frenk $\&$ White (1997) with f$=$0.01. 
The ``concentration'' of a halo, $c = r_V/r_s$, is linked to the 
characteristic density $\delta_{char}$ by means of the definition of the 
virial radius so that~:

\begin{equation}
\delta_{char}={200 \over 3} {c^{3} \over
\bigl[\ln(1+c)-c/(1+c)\bigr]}
\end{equation}

This implies that the ``circular velocity'' at radius $r$ is given by~:

\begin{equation}
\biggl({v_c(r)\over v_{200}}\biggr)^2={1 \over x}
{\ln(1+cx)-(cx) /(1+cx) 
\over  \ln(1+c)-c/(1+c)}
\end{equation}

\noindent where $v_{200} \equiv (GM/r_V)^{1/2}$ is the velocity at the 
virial radius, and $x=r/r_V$ is the radius in units of virial radius. 

\noindent The number density of halos that collapse at a given redshift/time 
and with a given mass is easily determined from a simple Press--Schechter (PS) 
prescription (Press $\&$ Schechter 1974).

\begin{eqnarray}
n(M,z)dM = & \sqrt {2 / \pi} (\rho_0 / M^2) (\delta_{0c}(1+z) /
\sigma_0(M)) \times \nonumber \\
 & |d\ln \sigma_0(M) / d \ln M| \times \nonumber \\
 & \exp - (\delta_{0c}(1+z)^2 / 2 \sigma_0(M)^2)~dM 
\end{eqnarray}

If the power spectrum can be approximated by a power law $P(k) \propto k^n$,
the low--mass end is $n(M)dM \propto M^{(n-9)/6} dM$. The CDM spectral index 
varies between $-3$ and 1, yielding a slope $\gamma =-2$ to $-1.67$,
close to $-2$ on galaxy scales. 

We intend hereafter to use a simple prescription
and focus on the astrophysics of galaxy formation and evolution. 
This will make the comparison with previous papers easier.
In order to implement the simpler PS formalism, crude
assumptions must be done. As a matter of fact, we need the formation rate of 
haloes versus time or redshift, while the time derivative of the PS number
density only gives the net creation rate of haloes with mass $M$~:

\begin{equation}
dn(M,z)/dt=dn_{for}(M,z)/dt-dn_{des}(M,z)/dt 
\end{equation}

The first term
of the right--hand part of the equation describes the formation of haloes 
with mass $M$ from haloes with masses $<M$, while the second term describes
the destruction rate to haloes with masses $>M$. For a given mass,
the formation rate first dominates ($dn(M,z)/dt>0$), then the destruction rate
dominates ($dn(M,z)/dt<0$). As {\it e.g.} in Haehnelt $\&$ Rees (1993),
we chose to take $dn_{for}(M,z)/dt=dn(M,z)/dt$ in 
the first case (destruction is neglected), and 0 in the second 
case (formation is neglected), to estimate the space density 
of newly formed haloes. This crude assumption leads to neglecting
the effect of galaxy merging which anyway seems to be weak.

\subsection{Gas cooling and dissipative collapse}\label{modelo_collapse}

Hierarchical theories of structure formation require a feedback mechanism 
in order to prevent most of the 
material from collapsing into sub-galactic objects at high redshifts 
({\it e.g.} Efstathiou 1992). So, we introduce a cut--off at low 
circular velocities and high redshifts preventing gas cooling if 
$v_{200} < 100$ km s$^{-1}$ and $2 < z_{coll} < 10$. As shown by 
Kauffmann {\it et al.} (1993), this does not alleviate the problem of the 
dwarfs.

The previously shock--heated baryon component starts 
cooling down in the potential wells of dark matter haloes and 
collapses into disks. The cooling time at a given 
halo radius is computed according to 
Fall $\&$ Rees (1985) and leads to the definition of a cooling radius 
($r_{cool}$) as a 
function of redshift. At this redshift, only gas inside the cooling radius 
(or the virial radius, in case the latter is smaller) cools and is 
available for star formation. Normally, the cooling process depends on gas 
metallicity but here we do not take this dependence into account, and we shall 
only consider the solar metallicity case.
If we assume angular momentum conservation, then the collapse of the baryonic 
matter stops when rotational equilibrium is reached. An exponential 
disk forms. To determine the final radius of the disk relatively to 
the initial radius of the halo, we apply the 
formalism proposed by Fall $\&$ Efstathiou (1980), and Mo, Mao $\&$ White 
(1998), under the standard assumption that the disk and halo
share the same specific angular momentum, even 
though numerical simulations (Navarro $\&$ White 1994, Navarro $\&$ Steinmetz 
1997) seem to hint that this may be too crude.
The typical length scale $r_0$ for the exponential 
disk is, again following Mo, Mao $\&$ White (1998)~:

\begin{equation}\label{eq:erre0}
r_0 \, = \, \lambda \, \min (r_V, r_{cool}) \, f(c,m_d,\lambda)
\end{equation}

\noindent where $f$ is a function of the dimensionless spin 
parameter $\lambda$, the halo 
concentration factor $c$, and the fraction $m_d$ of the halo mass that 
settles into the disk. A sufficiently accurate approximation (Mo, Mao $\&$ 
White 1998) to $f$ is~:

\begin{eqnarray}
f \approx &
\left({\lambda \over 0.1}\right)^{-0.06+2.71m_d+0.0047/\lambda} 
(1-3m_d+5.2 m_d^2) \times \nonumber \\
 & \times (1-0.019 c+0.00025 c^2+0.52/c), 
\end{eqnarray}

We note that only disks and dwarfs can form in this formalism. As we 
previously mentioned in section~\ref{introducao}, the 
formation of elliptical galaxies (and of bulges of spiral galaxies) has to be 
explained by the merging of disks. Kauffmann {\it et al.} (1994) and Cole 
{\it et al.} (1994) showed that this merging process can easily explain the 
current fraction of giant ellipticals among bright galaxies (about 10 \%). 

\begin{figure}[htbp]
\hspace{1.cm}
\epsfysize=9cm
\epsfbox{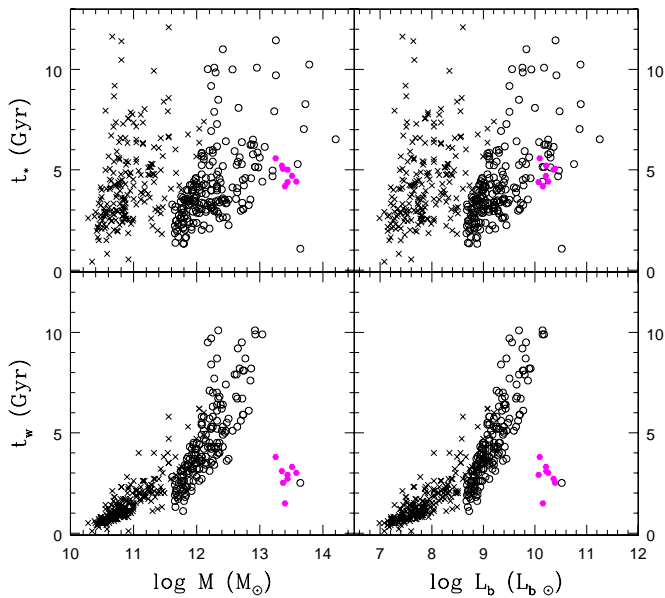}
\caption{The characteristic time scale for star formation $t_*$ and the age at 
which each galaxy has its first wind $t_w$ as a function both of the halo mass 
(left--hand panels) and blue luminosity (right--hand panels). The free 
parameters and symbols are the 
same as in figure~\ref{fig:expansao} except for the newly added filled 
circles, which account for a particular class of high luminosity, low surface 
brightness, galaxies (see text for details). Only galaxies that have winds
have been plotted in the bottom panels}
\label{fig:epocas}
\end{figure}

\subsection{Star formation rate and initial mass function}
\label{modelo_starform}

Stars begin to form with the gas available inside the cooling (or the 
virial) radius according to a given star formation rate $\psi(t)$. Here we 
shall consider a simple case of direct proportionality to the gas surface 
density $\Sigma_{gas}$, which seems to be supported by the latest 
observational evidence (Kennicutt 1998). We also introduce a critical 
threshold $\Sigma_c$, according to Toomre (1964) and Kennicutt (1989). We 
here adopt the numerical value determined by Kennicutt (1989) for an ensemble 
of disks~:

\begin{equation}
\Sigma_c (M_{\odot} pc^{-2}) \, \simeq \, 0.40 \, v_{200} (km s^{-1})/r_{disk} (kpc)
\end{equation}

\noindent knowing that the half-mass radius of an exponential disk is~:

\begin{equation}\label{eq:rdisk}
r_{disk} \, = \, 1.68 r_0
\end{equation}

\noindent This will prevent star formation from occurring whenever the gas 
surface density is below this limit. Otherwise we have~:

\begin{equation}\label{eq:sfr}
\psi(t) \, =  \, \frac{\Sigma_{gas}(t)}{t_*}
\end{equation}

\noindent with $t_*$ derived from the core dynamical times as~:

\begin{equation}\label{eq:testar}
t_* \, = \, \beta \, t_{dyn,disk}
\end{equation}

\noindent Here $\beta$ is a free parameter defining the efficiency of star 
formation per dynamical time,\footnote{Do notice, 
when comparing definitions, that our $\beta$ parameter here is the Guiderdoni 
{\it et al.} (1998) one divided by $2 \pi$.} and~:

\begin{equation}
t_{dyn,disk} \, = \, \frac{r_{disk}}{v_{200}}
\end{equation}

After their formation, stars are placed on the zero-age main sequence of 
the HR diagram according to an initial mass function. In this work we have 
chosen to describe the initial mass function by the standard form 
$\phi(m) \propto m^{-x}$ with slope $x$~=~1.7 
for stars more massive than 2 $M_{\odot}$ (Scalo 1986), the 
Salpeter slope $x$~=~1.35 for 
masses between 1 and 2 $M_{\odot}$, and $x$~=~0.25 below 1 $M_{\odot}$. 
A model of spectro-photometric evolution (Guiderdoni $\&$ Rocca-Volmerange 
1987, 1988 and Rocca-Volmerange $\&$ Guiderdoni 1988) issued for a burst of 
0.1~Gyr duration with upgraded tracks from Schaller {\it et al.} (1992) and 
Charbonnel {\it et al.} (1996) is used to compute the mass/luminosity ratios 
and the colors of each generation of stars according to their age. Then the 
burst evolution is convolved with the star formation rate history of each 
galaxy.

\subsection{Feedback processes}\label{modelo_feedback}

The mechanism of star formation itself is regulated by feedback due to the 
explosion of massive stars. The standard modelling of feedback due to 
supernovae in semi--analytic models of galaxy formation is the following:
supernova--driven winds expel gas from the galaxy and, as 
a consequence, less star formation occurs (Larson 1974, Dekel $\&$ Silk 1986,
Bressan {\it et al.} 1994). 
To determine the time $t_w$ when the galaxy loses its gas mass, we keep track 
of the balance between the total supernova input energy, responsible for the 
heating of the inter-stellar medium, and the energy 
required to eject the gas from the radius $r$ of a galaxy at time $t_w$~:

\begin{equation}\label{eq:balance_lacey}
\frac{1}{2} \, M_{gas}(t_w) \, v_{esc}(r)^2 \, = \, \epsilon \, 
\int_0^{t_w} \, E_{SN} \, \frac{dN_{SNII}(t)}{dt} \, dt
\end{equation}

considering that~:

\begin{equation}
\int_0^{tw} \, \frac{dN_{SNII}(t)}{dt} \, dt \, = \, \eta_{SN} \, M_*(tw)
\end{equation}

\noindent The gas escape velocity $v_{esc}(r)$ at radius $r$ is computable
for the NFW profile~:

\begin{equation}
\biggl({v_{esc}(r)\over v_{200}}\biggr)^2={2 \over x}
{\ln(1+cx)-(cx) /(1+c) 
\over  \ln(1+c)-c/(1+c)}
\end{equation}

It is estimated at the half--mass radius $r_{disk}$.
Here $E_{SN}$ is the input energy per supernova, and $\eta_{SN}$ is the 
number of supernovae per unit mass of formed stars that depends on the 
initial mass function. We adopt the standard values $E_{SN} = 10^{51}$ erg,
and $\eta_{SN} = 4 \times 10^{-3}$ $M_{\odot}^{-1}$ (computed for our 
assumed IMF). The Type II/Ib supernovae 
formation rate  $dN_{SNII}(t)/dt$ is computed from the star formation rate 
history. All models deliberately neglect the contribution from Type Ia. 
A fine--tuning ``efficiency factor'' $0 \leq \epsilon \leq 1$ is 
generically introduced to account for the fraction of the energy that is 
radiated away and lost by the system. Numerical simulations 
(Thornton {\it et al.} 1998) seem to privilege $\epsilon \sim 0.1$. In fact,
the efficiency parameter could be still smaller in disk galaxies because 
of the geometry that allows supernova bubbles to blow out. In 
section~\ref{fit_parametros_epsilon} 
we will analyze and discuss the behavior of our model with this 
parameter which is always used as a ``fudge factor''. 

\subsection{Expansion of the stellar disk}

When a galaxy loses mass, this process can occur either impulsively - when the 
material is ejected within a time scale that is short compared to the 
dynamical time of the system --, or adiabatically -- when mass 
loss proceeds gradually. In our models we have 
chosen to consider the first approach only, which seems the most realistic. 
This implies that the velocities of the stars are not affected during the 
whole process and their velocity dispersion immediately after the 
(impulsive) mass loss is the same as it was before. A straightforward 
consequence of mass loss is a shallowing of the system's potential.
Stars will start describing larger orbits. This translates into a
radial expansion of the stellar 
material of the disk, that we treat here according to Hills (1980). Note that 
this extension of the galactic radius is crucial for surface brightness 
considerations as it leads to surface brightness dimming. The final radius of 
the stellar material, which will allow us to determine the surface brightness 
of each galaxy after mass loss, is related to the initial radius by equation~:

\begin{equation}\label{eq:rayexpansion}
\frac{r_f}{r_i} \, = \, \frac{M_f}{2 \, \left[\frac{M_i}{2} \, - \, 
\left(M_i-M_f \right) \right]}
\end{equation}

\noindent where the $i$ subscript stands for initial values and $f$ for final 
ones, after mass loss has occurred. All variables of this equation have been 
defined at half--mass radius, and $r_i$ refers to the initial stellar radius 
obtained from the halo radius by means of the (dimensionless) spin parameter 
(see equations~\ref{eq:rdisk} and~\ref{eq:erre0}). 

\begin{figure*}[htpb]
\centering
\psfig{figure=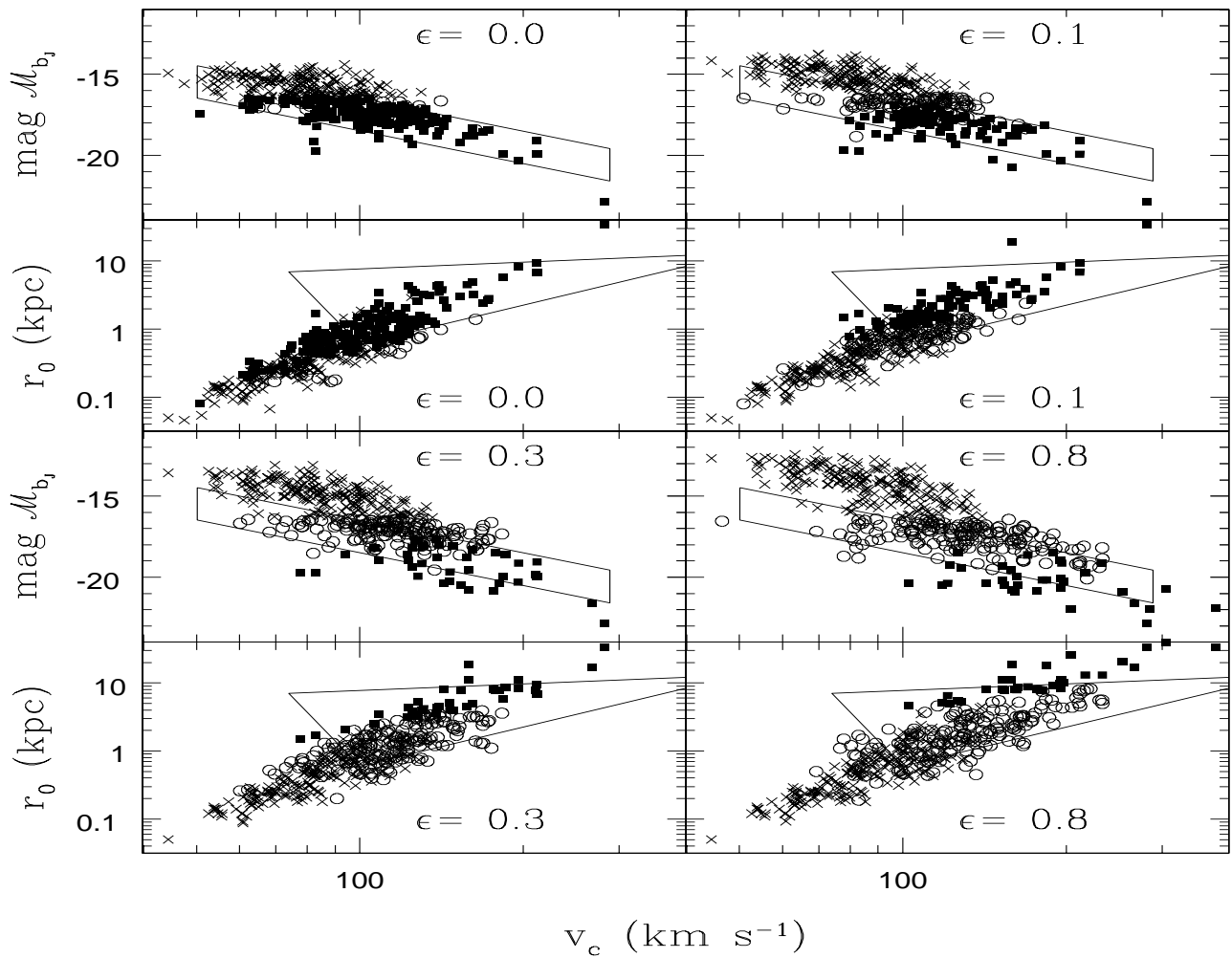,height=14.cm,width=18.cm}
\caption{Tully--Fisher relation (upper panels for each $\epsilon$) 
and typical scale length as a function of circular velocity (lower panels 
for each $\epsilon$) for model galaxies. Boxes 
indicate observational values reported by Kraan-Korteweg, Cameron $\&$ 
Tammann (1988; upper panels), and Courteau (1996, 1997; lower panels). 
The wind efficiency coefficient $\epsilon$ is taken equal to (a) 0, (b) 0.1, 
(c) 0.3, and (d) 0.8. Symbols 
are the following~: crosses note objects that do not pass the flux and surface 
brightness threshold criteria, while objects that fulfill these constraints 
are represented both by circles - objects with $B - V > 0.85$ - or by squares 
- bluer $B - V \leq 0.85$ galaxies.}
\label{fig:tullyfisher}
\end{figure*}

\subsection{Surface--brightness threshold}

Finally, we shall introduce both a flux limit and a surface--brightness 
threshold 
that simulate observational limits and we analyze the behavior and 
sensitivity of our model predictions to these effects 
(section~\ref{funcoes_luminosidade}). In fact, it is often the case that the 
isophote taken to compute the surface brightness limit of a given survey may 
(1) differ slightly from galaxy to galaxy (if it depends on the magnitude 
or the size of the object); 
(2) depend on selection criteria (if the filter used for 
target selection is not the same as the one used to compute the 
luminosity function); (3) vary for different images of one same field taken in 
different nights and/or under different observational conditions. These 
effects can lead to differences of up to 1 magnitude
(E. Bertin, private communication).
As a reasonably fair approximation, the surface--brightness 
limit imposed in our models is computed at the effective radius $r_e$.
Of course, surface--brighness effects are likely to be more subtle than a 
simple surface--brightness ``cut--off''. But little more can be done from 
the description of the observations provided in the literature. 

\subsection{Monte--Carlo realizations}

We present in this section some of the results of the model described above 
for output redshift $z=0$, while emphasizing the surface brightness effects 
that we want to explore (presented in section~\ref{introducao}). One could 
represent these results by a mean value plus $1\sigma$ ``error bars'' but, 
because of the non gaussianity of the distributions, this would be somewhat 
misleading. So, we preferred to draw Monte--Carlo realizations of the model. 
Thus, in the set of figures that follow, we represent a given number of 
objects (realizations), separating them accordingly to a magnitude and surface 
brightness criterion established by observational limits. 

Both classes of objects are 
represented in equal numbers (e.g. 200 of each). 
Unless stated otherwise, we have adopted 
limits $\mu_{b_J}=24$ mag arcsec$^{-2}$ and absolute magnitude 
$\cal{M}$$_{b_J}= - 16.5$ in agreement with those existing in a typical 
field galaxy 
redshift survey such as the Stromlo-APM (see Loveday {\it et al.} 1992). 
Likewise, we draw as many galaxies as we need so that those passing the 
surface brightness and flux 
criteria will approximately match the number of galaxies used by Loveday 
{\it et al.} (1992) for their luminosity function determination. 
In the next set of figures, we will fix $\epsilon = 0.8$ and $\beta = 400$, 
unless explicitly stated otherwise. 
The choice of these values will be justified later, in 
sections~\ref{fit_parametros_epsilon} and \ref{fit_parametros_beta}.

\begin{figure}[htbp]
\centering
\vbox{
  \subfigure[$\beta$~=~400]{ 
\psfig{figure=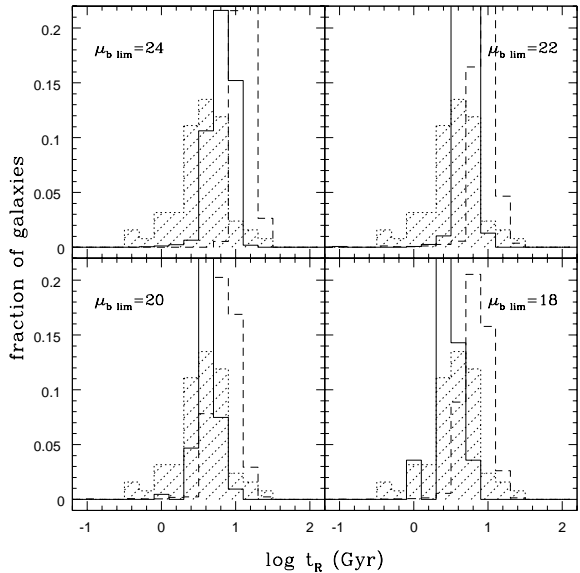,height=8.cm}}
  \subfigure[$\beta$~=~100]{ 
\psfig{figure=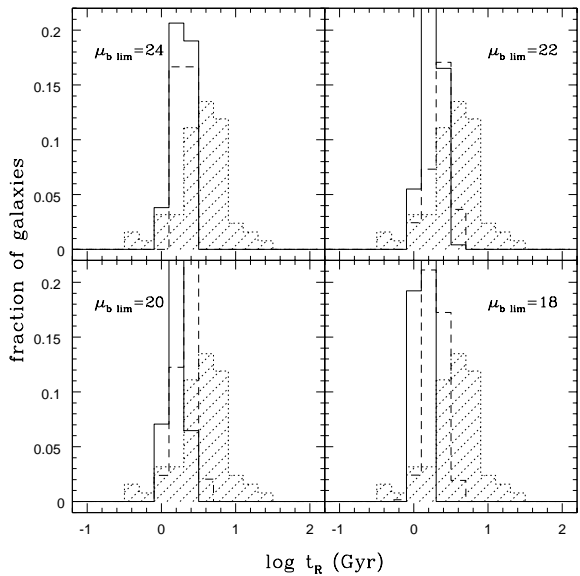,height=8.cm}}
      }
\caption{Comparison between the observed distribution for the Roberts time 
(by Kennicutt {\it et al.} 1994) -- hatched histogram -- with distributions 
issued by our 
model for two different values of $\beta$, given below each set of figures. 
The histogram drawn with a continuous line is for objects that pass both 
magnitude and surface brightness thresholds, whereas fainter objects that 
fail to pass these criteria are represented by the dashed--line histogram. 
The surface brightness limit (in $b_J$ magnitudes arcsec$^{-2}$ at effective 
radius) is indicated for each case. The $\epsilon$ parameter is fixed at 0.1.}
\label{fig:allrobertstime}
\end{figure}

\section{Sensitivity to the parameters}\label{fit_parametros}

\subsection{Radial expansion}\label{spin_plus_epocas}

Figure~\ref{fig:expansao} shows quite straightforwardly the lowering of 
surface brightness~: as mass loss due to
supernova--driven winds occurs, subsequent radial 
expansion naturally causes dimming. The objects that undergo larger
radial expansion and higher mass loss are mainly the  ``dwarfs''
(with small size and mass). 
However, there is also a fraction of galaxies with higher masses 
that become ``invisible'' (through a significant radial expansion). 
These could be the ``Malin~1 type'' extended 
disks with faint surface brightnesses that we were referring to previously.

Note that if we were to enhance the strength and frequency of winds (by 
adopting higher $\epsilon$ values) during a galaxy's lifetime, the radial 
expansion would be more pronounced and final radii could reach up to  
20 kpc. For the same reason, the $\mu_{be}$ relation (lower panel of 
figure~\ref{fig:expansao}) would also appear more scattered.

Figure~\ref{fig:epocas} shows different time scales against both halo mass 
and blue luminosity. 
In general, there does seem to be a value of the halo mass that acts as a limit
for ``visibility''~: 
low-mass objects (say below $10^{11.5}$ $M_{\odot}$) are not 
bright enough (in magnitude nor in surface brightness) to be ''observed''.
However, the representation of figure~\ref{fig:epocas} leaves out 
the most massive and high luminosity but low surface brightness objects such 
as those of the type of Malin~1 with $L_B \simeq 1.9 \times 10^{11} 
L_{B \odot}$, $\mu_{e} = 22.2 \pm 0.2$ V--mag/arcsec$^2$ (Bothun {\it et al.} 
1987), but it does show (as filled circles) a less extreme class of objects~: 
the typical giant low surface brightness spiral galaxies observed by 
Sprayberry {\it et al.} (1995) with 
$1.06 \times 10^{10} \leq L_B \leq 2.3 \times 10^{11} L_{B \odot}$, 
$21.17 \la \mu_{e} \la 26.32$ $B$--mag arcsec$^{-2}$. 
In fact, this type of galaxies can be considered to be somewhat 
rare (from observations) and this seems to be supported by the results of 
our model, which predicts that, for $L_B \ga 10^{10} L_{B \odot}$,
the number density of low surface--brighness objects
(with $\mu_B \geq 24$ mag arcsec$^{-2}$) is only $\sim 1\%$ of the 
population with a high surface brightness.
Finally, one should note that some galaxies 
suffer no wind at all during their entire lifetime (e.g. 
$\sim 10\%$ for ``visible'' ones). The objects with wind are 
plotted in the bottom panels of figure~\ref{fig:epocas}. 

\begin{figure*}[htpb]
\centering
\hbox{
  \subfigure[$\epsilon$~=~0.3]{ 
\psfig{figure=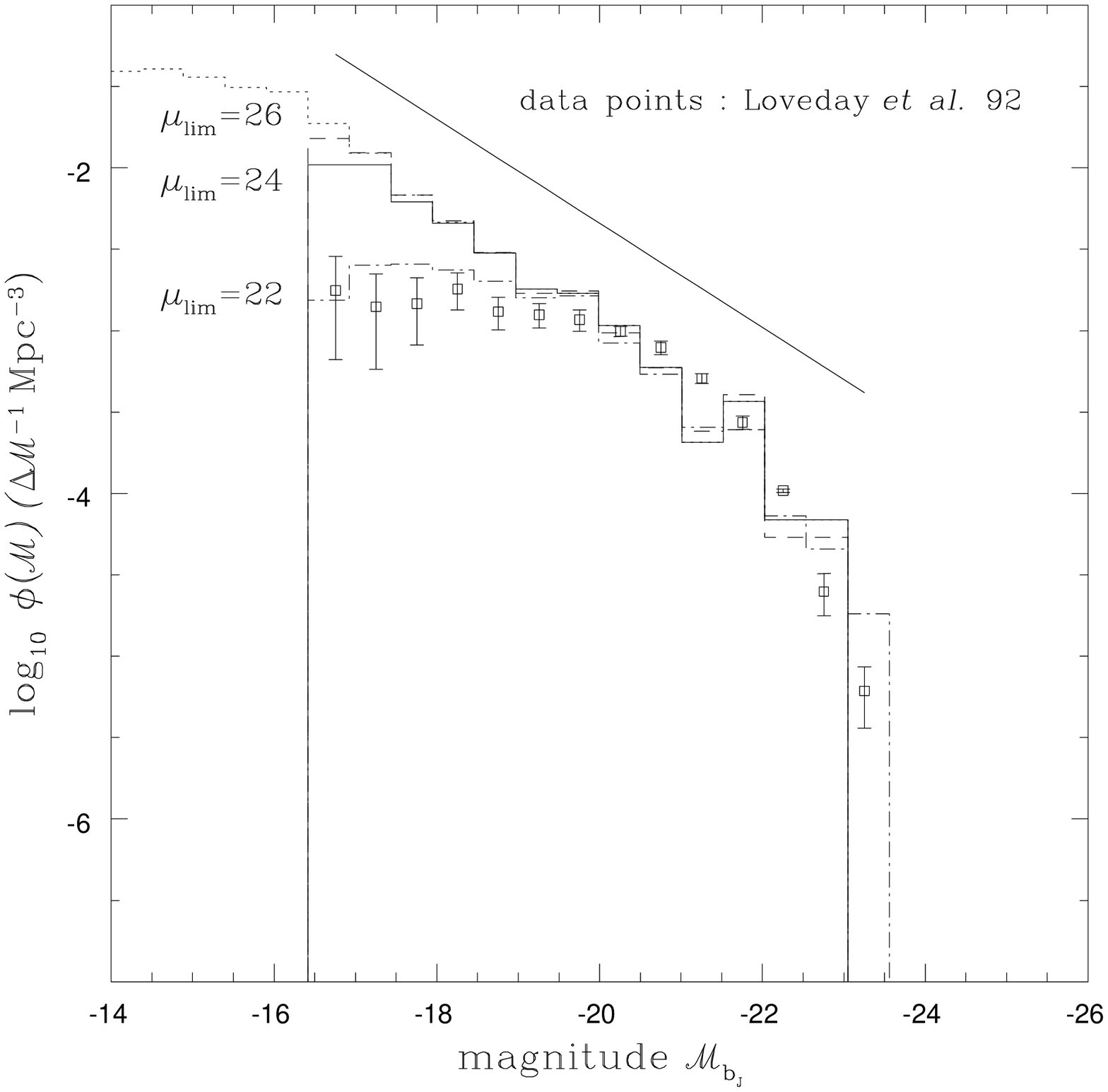,height=9.cm}}
  \subfigure[$\epsilon$~=~0.8]{ 
\psfig{figure=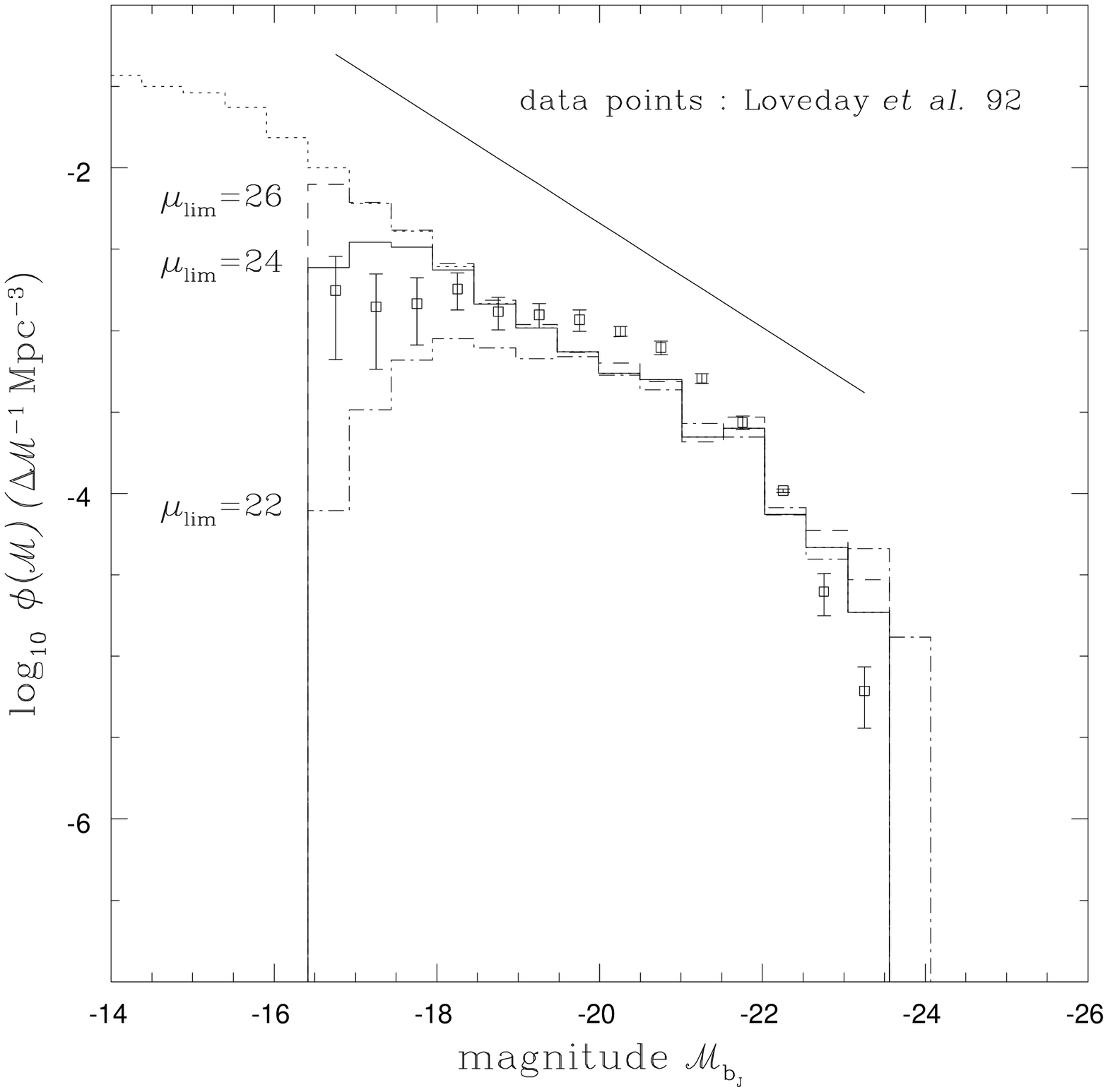,height=9.cm}}
      }
\caption{Luminosity function normalized per Mpc$^3$ and per magnitude bin in 
the $b_J$ band. Observations (by Loveday {\it et al.} 1992; data points) have 
been superimposed to the models (histograms) for (a) $\epsilon = 0.3$, 
(b) $\epsilon = 0.8$. The histograms with the dot--dashes, solid line and 
dashes indicate model galaxies that pass the survey 
flux limit, as well as the surface brightness limit that is indicated near 
each histogram for the respective mock luminosity function. The number of 
galaxies is approximately equal to the value reported for the observations. 
The histogram with the dotted line shows 
the model galaxy distribution as issued before submission to these selection 
criteria (that is for all galaxies, none excluded). The slope 
$\alpha = -1.8$ of the Coma cluster luminosity function as determined by 
Lobo {\it et al.} (1997) is also represented (straight line).}
\label{fig:difSBlim}
\end{figure*}

\subsection{Feedback efficiency $\epsilon$}\label{fit_parametros_epsilon}

Figure~\ref{fig:tullyfisher} shows both the Tully--Fisher relation (TF) and 
the disk size versus circular velocity relation adopted for typical 
late--type galaxies. 
We display four cases of our model results~: no wind, $\epsilon=0$;
an $\epsilon = 0.1$ low--wind efficiency scenario; an intermediate case 
$\epsilon = 0.3$; and an $\epsilon = 0.8$ strong--wind possibility. 
In each figure we have 
limited by boxes the regions corresponding to the values observed by, 
respectively, Kraan-Korteweg, Cameron $\&$ Tammann (1988) for the TF relation 
of spirals (notice that their data are consistent with $H_0=57$ 
km$^{-1}$ Mpc$^{-1}$ and are rescaled to our $H_0=50$ km$^{-1}$ Mpc$^{-1}$ 
adopted value), and Courteau (1996, 1997) for 
the disk--velocity relation for a sample of nearby normal spirals. In this set 
of plots we separate model galaxies 
by colors, setting the border at $B - V = 0.85$, which is the typical color
of S0 galaxies at $z=0$ (Fukugita {\it et al.} 1995).

We see from figure~\ref{fig:tullyfisher} that the $\epsilon = 0$ model is 
the one that reproduces best the general trend and dispersion of the 
observations for large disks, in spite of a theoretical scatter which seems 
to be too large. We would thus tend to prefer this value although up to 
$\epsilon = 0.1$ we still get a crude agreement of the model with the 
observations. It is well--known that the theoretical fit to these 
relations is not straightforward. In particular, 
the steep luminosity function that theoretical models produce
in the standard CDM is intimately 
linked to the difficulty to fit the Tully--Fisher relation with such models 
(Kauffmann {\it et al.} 1994; Cole {\it et al.} 1994;
Somerville $\&$ Primack 1998). But, in spite of 
these shortcomings, we do seem to manage to obtain a generally good relation 
between the 
dynamics and the photometry, as is patent in this ensemble of figures~: we 
have the correct galactic luminosities inside the halos, and these galaxies 
have the correct size (radii). This last point is crucial for surface 
brightness considerations once we discuss the predictions for the luminosity 
functions.

\subsection{Star formation efficiency}\label{fit_parametros_beta}

The Roberts time (Kennicutt {\it et al.} 1994) gives an indication of the 
future star formation time scale of a galaxy at time $t_0$~:

\begin{equation}
t_R(t_0) \, = \, \frac{M_{gas}(t_0)}{\psi(t_0)}
\end{equation}

\noindent $\psi(t_0)$ here is either given by equation~\ref{eq:sfr} 
(with $t = t_0$) or null if 
the Toomre instability criterium (Toomre 1964) is not verified, that is, if 
the gas surface density does not overcome $\Sigma_c$ (see 
section~\ref{modelo_starform}). 

We thus computed the Roberts time for a set of model galaxies and compared its 
distribution with that observed by Kennicutt {\it et al.} (1994) for a sample 
of about 60 disk galaxies. The value of the $\epsilon$ 
parameter was left constant at 0.1, as a result from 
section~\ref{fit_parametros_epsilon}. This procedure provided us with 
the best value for the constant 
$\beta$, which we finally fixed at 400 -- see figure~\ref{fig:allrobertstime}.
Because we do not know the surface brightness threshold for the Kennicutt 
{\it et al.} data, we cannot determine a unique possibility for the 
fit, and consequently for $\beta$. But the degeneracy thus created is not 
limitating, as is illustrated in each panel of 
figure~\ref{fig:allrobertstime}a, where different realizations of the model 
for different surface brightness thresholds -- but the same value for $\beta$ 
-- are compared to the observed data. 
The distribution of Roberts times does not seem to depend too 
much on the surface brightness threshold. 
The Kennicutt {\it et al.} limit in surface brightness seems to lie around 
20 $-$ 22 $b_J$~mag arcsec$^{-2}$, as estimated for $\beta$ = 400. 

\begin{figure*}[htpb]
\hspace{1.cm}
\epsfysize=16cm
\epsfbox{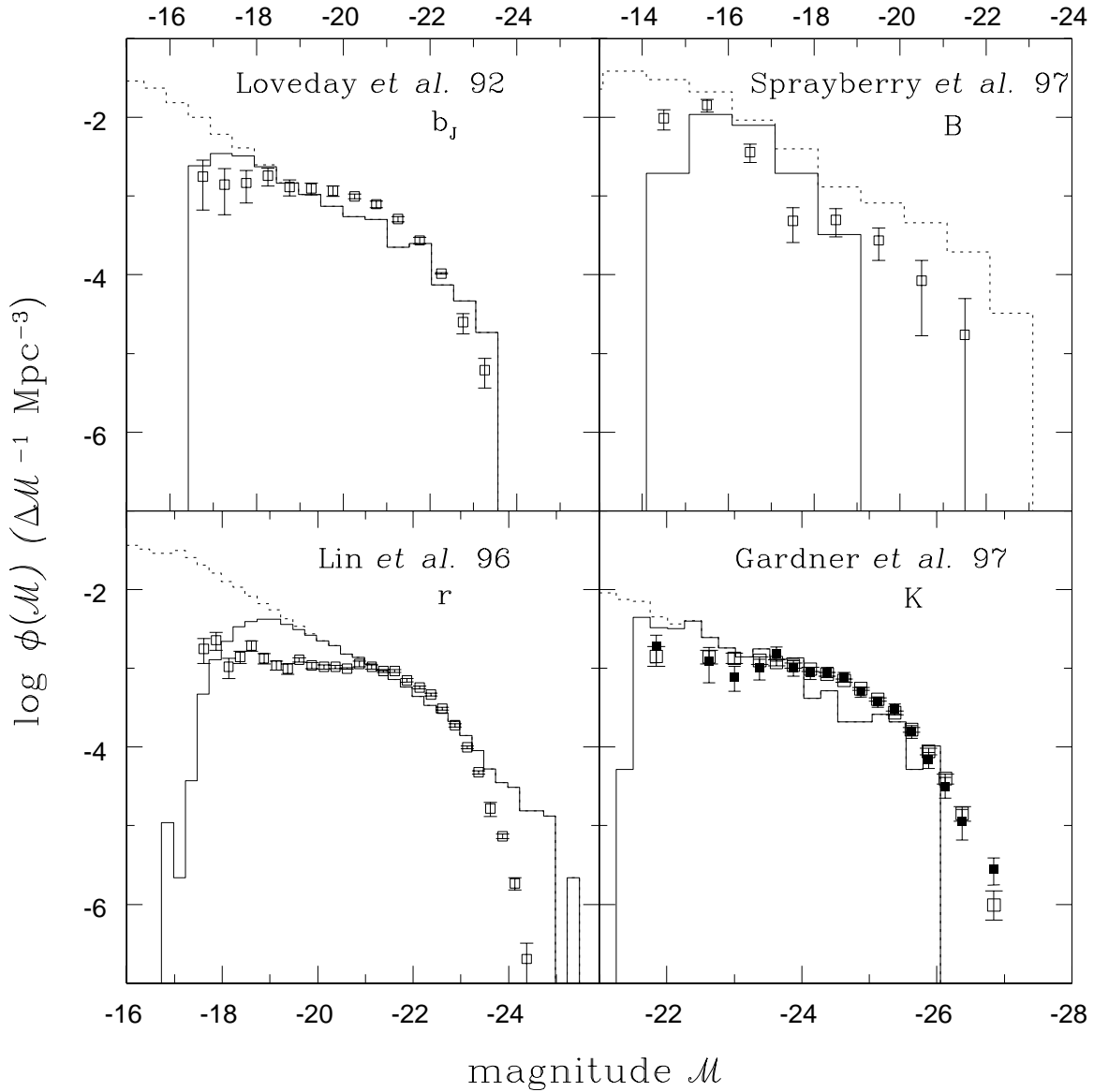}
\caption{Same figure as~\ref{fig:difSBlim}b ($\epsilon=0.8$) for four 
different photometric 
wavebands. Observations (data points) available in each filter are also 
displayed for comparison. Surface brightness and flux limiting criteria 
applied to the simulations are approximately the ones reported by each 
survey (see text).}
\label{fig:4fdls}
\end{figure*}

\section{Predictions for the luminosity functions in the local Universe}
\label{funcoes_luminosidade}

Galaxy luminosities computed with the model allow us to build synthetic 
luminosity functions at low redshift and in different photometric bands. These 
will be compared with observational results in this section.

\subsection{Effect of $\epsilon$}

In figure~\ref{fig:difSBlim}, the normalized luminosity function per Mpc$^3$ 
and per magnitude interval, as issued by the model, is compared with 
observations for field and cluster galaxies.

We assume that all objects fainter than a given surface brightness detection 
limit will not be observed. The figure shows that, for bright galaxies, this 
threshold is not constraining, as one would expect, but the faint part of the 
luminosity function is significantly modified for different surface brightness 
limits. This change is translated into a flattening of the histogram that gets 
more pronounced as the critical threshold gets more severe. Data points from 
Loveday {\it et al.} (1992) are also displayed in the same figure for 
comparison. 

It is well--known that, with the standard CDM model, the simultaneous fit of 
the Tully--Fisher relation and luminosity function is difficult, because 
there are too many haloes (see e.g. Kauffmann {\it et al.} 1993). This is
still true when surface--brightness effects are taken into account, 
with the crude modelling of supernova feedback.
Even if the $\epsilon = 0.3$ model, which is already inconsistent with TF, 
accompanies best the part of the 
observational data around $\cal{M}$$^{\star}$ (panel a), it fails to reproduce 
the faint end slope with the surface brightness 
limit reported by Loveday and collaborators for their survey (around 24.5 
$b_J$ mag arcsec$^{-2}$). Setting $\epsilon = 0.8$ allows a 
much better description for the ``dwarf'' population as 
figure~\ref{fig:difSBlim}b reveals. As a matter of fact, results are highly 
sensitive to the value of $\epsilon$ and this work might hint that 
$\epsilon$ changes along the luminosity sequence, but 
this simply illustrates that modelling galactic winds is quite a 
complex task, not yet fully accomplished with a single free parameter.
Still, we shall consider that 
$\epsilon = 0.8$ gives a very reasonable and general representation of the 
faint end of field galaxy luminosity functions.

Figure~\ref{fig:difSBlim} also shows the slope $\alpha = -1.8$ of 
the Coma cluster luminosity function as determined by Lobo {\it et al.} (1997) 
in the V band up to magnitude $\cal{M}$$_V \simeq -14.5$ with limiting central 
surface brightness estimated as $\mu_V (0) = 24.5$ mag arcsec$^{-2}$.
Note that, if we take $B - V = 1$ as a mean colour index for Coma 
galaxies (mainly ellipticals), and the simple 
relation connecting the surface brightness at effective radius to the central 
surface value, that is $\mu_{e} = \mu(0) + 1.127$, we get 
$\mu_{b_J} \sim 26.4$ mag arcsec$^{-2}$
for the practical threshold. The normalisation for the 
Coma data is arbitrary. It is true that we can only take this slope as 
indicative, as it was determined in a different waveband. Nevertheless, the 
similarity of it with the shape of the $\mu_{lim} =26$ mag arcsec$^{-2}$
model histogram is quite striking, for $\epsilon = 0.3$ as well as
$\epsilon = 0.8$.

\subsection{Effect of $\mu$ within a given filter and looking at different 
wavebands}

Given all these considerations, we will henceforth fix the $\epsilon$ 
parameter to 0.8, to reproduce the faint--end of the luminosity function.
Figure~\ref{fig:4fdls} gives the model results convolved with four different 
filters. For each simulation we have applied flux and surface brightness 
limits corresponding to what was indicated by the authors of the respective 
observations. Thus, we have adopted the following thresholds~: 
$\cal{M}$$_{b_J} = -16.5$ 
and $\mu_{b_J}=24$ mag arcsec$^{-2}$ for Loveday {\it et al.} (1992) at 
$z = 0$; $-14 \le$ $\cal{M}$$_B \le -22$ and $23.1 \le \mu_B \le 26.1$ for the 
deeper observations of Sprayberry {\it et al.} (1997), still in a blue 
waveband and at $z = 0$; $\cal{M}$$_K = -21.5$ and $\mu_I = 22$ mag 
arcsec$^{-2}$ for the K-band survey (selected in I) of Gardner {\it et al.} 
(1997) at mean redshift $z = 0.14$; and $\cal{M}$$_r = -17.5$ and 
$\mu_r = 22.2$ mag arcsec$^{-2}$ for the Lin {\it et al.} (1996) 
$r_{Gunn}$-type filter at 
mean redshift $z = 0.1$. Do notice that these values are indicative and a 
margin of $\pm 0.5$ mag arcsec$^{-2}$ is allowed by the robustness of 
our models, as can be seen in figure~\ref{fig:difSBlim}. This should give us 
some margin to deal with uncertainties in the values estimated by the authors 
and with our choice of approximating the threshold to the value at effective 
radius. In each case, simulations were performed as many times as necessary 
to draw the same number of ``observable'' galaxies as the one reported 
by the respective surveys (1769 galaxies for Loveday {\it et al.} 1992; 693 
galaxies for Sprayberry {\it et al.} 1997; 510 galaxies for Gardner 
{\it et al.} 1997; and 18678 for the Las Campanas survey of Lin {\it et al.} 
1996). In doing so, we get Poisson scatters combined with intrinsic features 
of the luminosity functions that compare to the observational features and 
error bars.

The overall match at the faint end between data and simulations taking 
into account observational biases is quite remarkable, given the crudeness 
of our assumptions in this heuristic model. 
The rough consistency obtained for different wavebands probably unveils the 
presence of the same underlying galaxy population, despite the 
fact that different filters preferentially sample different stellar 
populations.

\section{Summary and Conclusions}\label{conclusoes}

This heuristic work shows how observational selection effects -- 
especially surface brightness thresholds that are mainly present in field 
surveys -- could affect the predictions of semi--analytic models of galaxy 
formation and evolution. In particular, we illustrate the importance 
of these biases in the determination of the TF relation and luminosity 
functions, and we confirm that they do seem to contribute to the apparent 
mismatch between current models of galaxy formation and a host of observations.

With our simple model, we show that, while the star formation efficiency 
parameter $\beta$ can be fixed in a robust way, that is, 
independently of the adopted surface brightness limit, the same is no longer 
true for the wind efficiency parameter $\epsilon$. In fact, we have shown the 
strong sensitivity of the results to the unknown galactic wind intensity. 
Once we adopt the standard one--parameter model for the feedback, 
we find that 
this parameter is likely to vary along the luminosity sequence. As a first 
guess, we suggest taking $\epsilon \le 0.1$ for bright disks (according to the 
TF results, and loose theoretical considerations) and $\epsilon = 0.8$ 
to match the faint end of the luminosity functions, that is, 
for field ``dwarfs'' and for the faint galaxy population, both in the field 
as well as in clusters. 
Further research needs to be done in order to 
reach a more refined description of this mechanism and improve theoretical 
models, especially in their interplay with surface brightness effects. In 
particular, simulations of observations are necessary to take into account the 
effects of surface brightness that are more subtle than a simple ``cut--off'.

The main point of the paper is the following~: The standard 
modelling of supernova 
feedback via mass loss due to galactic winds is highly uncertain. Taking 
into account surface--brightness effects in a simple way already 
reveals the strong sensitivity of the results to the feedback.
Still, the results are in reasonable agreement with the observations,
The crude fit of the data with this type of theoretical luminosity functions 
-- at different surface brightness thresholds and at different wavelengths -- 
in spite of the uncertainties of the model, appears rather puzzling. It 
suggests that the discrepant waveband luminosity functions probably unveil the 
same underlying galaxy population. 

Clearly more work has to be dedicated to this issue. A realistic feedback 
theory is still needed for a robust modelling of faint galaxies in the 
semi--analytic approach. In any case, the introduction of 
surface--brightness biases into theoretical predictions, and more 
specifically into the fashionable semi--analytic models of galaxy formation and
evolution, is highly desirable, no less than is the careful estimate and
description of the actual surface brightness threshold by observers.

\begin{acknowledgements}
We are very grateful to J. Gardner for kindly providing us with the K-band 
luminosity function table data from Gardner {\it et al.} (1997) prior to 
publication. 
CL greatly thanks E. Bertin for the very helpful discussions. Finally,
we are grateful to an anonymous referee for his comments that greatly helped 
us improve the focus and presentation of this paper. During this work 
CL was supported by the fellowship reference BD/2772/93RM granted by JNICT, 
Portugal.
\end{acknowledgements}

\end{document}